\documentclass[12pt,letterpaper,fleqn]{article}           

\usepackage[top=1in,bottom=1in,left=1in,right=1in]{geometry}                                     

\usepackage{physics}

\usepackage{fullpage}                                     
\usepackage[doublespacing]{setspace}                                     
\usepackage{pdflscape}                                    

\usepackage{adjustbox}
\usepackage{booktabs}
\usepackage{makecell}
\usepackage{threeparttable}
\usepackage{graphics}
\usepackage{diagbox}

\usepackage{pdfpages}

\usepackage{caption}
\usepackage{varwidth}
\DeclareCaptionFormat{myformat}{%
  \begin{varwidth}{\linewidth}%
    \centering
    #1#2#3%
  \end{varwidth}%
}

\usepackage[latin9]{inputenc}                             
\usepackage[T1]{fontenc}                                  
\usepackage{lmodern}                                      
\usepackage{textcomp}                                     
\usepackage{amsmath}                                      
\usepackage{amsthm}                                       
\usepackage{amsfonts}                                     
\usepackage{amssymb}                                      

\usepackage[mathlines]{lineno}

\usepackage{xcolor}                                       
\definecolor{darkblue}{rgb}{0.0,0.0,0.66}                 
\usepackage[hyperfootnotes=false,bookmarksopen]{hyperref} 
\hypersetup{                                              
    pdffitwindow=false,                                   
    pdfstartview={XYZ null null 1.00},                    
    pdfnewwindow=true,                                    
    colorlinks=true,                                      
    linkcolor=darkblue,                                   
    citecolor=darkblue,                                   
    urlcolor=darkblue  }                                  

\usepackage{graphicx}                                     
\usepackage[position=bottom]{subfig}                      
\usepackage[section]{placeins}                            
\usepackage{float}                                        
\usepackage[justification=centering]{caption}             
\usepackage{tikz}                                         
\usetikzlibrary{decorations}                              
\usepackage{siunitx}

\usepackage{appendix}

\usepackage{titletoc}
\titlecontents{part}
[0pt] 
{} 
{\huge\bfseries\centering} 
{} 
{} 

\usepackage{natbib}                            

\usepackage{authblk}                                      

\usepackage{etoc}

\begin{document}

\begin{titlepage}

\title{Maximum Hallucination Standards for Domain-Specific Large Language Models}

\renewcommand\Affilfont{\small}
\renewcommand\Authsep{\hspace{0.7cm}}
\renewcommand{\Authands}{\hspace{0.7cm}}
\author{Tingmingke Lu\thanks{Department of Economics, WU (Vienna University of Economics and Business), tingmingke.lu@wu.ac.at}}

\date{\today}
\maketitle
\thispagestyle{empty}

\vspace{-.5cm}
\begin{abstract}
\doublespacing

Large language models (LLMs) often generate inaccurate yet credible-sounding content, known as hallucinations. This inherent feature of LLMs poses significant risks, especially in critical domains. I analyze LLMs as a new class of engineering products, treating hallucinations as a product attribute. I demonstrate that, in the presence of imperfect awareness of LLM hallucinations and misinformation externalities, net welfare improves when the maximum acceptable level of LLM hallucinations is designed to vary with two domain-specific factors: the willingness to pay for reduced LLM hallucinations and the marginal damage associated with misinformation. (JEL D8, K2, L5)

\end{abstract}

\end{titlepage}


\section{Introduction}
Large language models (LLMs) are being incorporated into our common workflows at a rapid pace \citep{NBERw32966}. However, LLMs are prone to hallucinations \citep{bender2021dangers}. LLM hallucinations can lead not only to counterproductive outcomes but also to severe consequences for users \citep[e.g.,][]{chatgptlawyer2023}. Moreover, LLMs are capable of generating false but credible-sounding contents at scale \citep{augenstein2024factuality}. Therefore, LLM hallucinations are likely to produce misinformation externalities, which are difficult to track and mitigate, posing significant risks \citep[e.g.,][]{bbcappleheadline}.

Existing research on LLM regulation has largely focused on areas such as market competition, data privacy, copyright, and ethical concerns.\footnote{See \citet{comunale2024economic} for a survey on LLM regulations.} As LLMs become widely adopted, the problem of hallucinations stands out as a critical yet under-addressed challenge. This paper provides the first analysis of regulating LLM hallucinations, filling a crucial gap in the current research. 

This paper studies the design of standards for domain-specific LLMs developed to carry out specialized tasks such as healthcare and legal services \citep[e.g.,][]{singhal2023large,chalkidis-etal-2020-legal}. Compared to general-purpose LLMs, these domain-specific LLMs are more deeply integrated into daily life because they are a more practical and pervasive technology for everyday applications. Considering LLMs as a new class of engineering products, I treat the LLM hallucination tendency as a product attribute in a simple model of demand for differentiated products.\footnote{Computer science researchers have developed various metrics to assess the severity of the LLM hallucination problem. For example, some researchers track LLMs' performance on common benchmarks over time, while others examine how frequently an LLM confabulates facts when summarizing a given document. Additionally, a Hallucination Vulnerability Index could be used to sort LLM hallucinations into six categories and three degrees of severity. See \citet{jones2025ai} for further details.} 

Just as accounting for consumer misperceptions of energy costs is important when designing minimum efficiency standards that address pollution externalities from energy-using durables, LLM regulations should also account for the fact that users may not be fully aware of LLM hallucinations. It is difficult, even for expert users, to identify such hallucinations. For instance, LLMs have limitations regarding ``long-tail'' knowledge because such rare, niche, or highly specialized facts are not well represented in their training data \citep{kandpal2023large}. When it comes to obscure or cutting-edge knowledge, even those with expertise in the broader domain can be vulnerable to incorrect but confident LLM outputs. 

Although hallucinations in LLMs are an inherent feature of these systems and cannot be completely eliminated, emerging techniques can reduce their occurrence \citep{jones2025ai}. I demonstrate that the optimal level of domain-specific LLM hallucination occurs when the marginal cost of lowering hallucinations equals the sum of consumers' domain-specific willingness to pay for reduced LLM hallucinations plus the domain-specific marginal misinformation damage. Moreover, I show that optimal domain-specific hallucination mandates accommodate users' imperfect awareness of LLM hallucinations, regardless of whether this imperfect awareness varies within each domain of use or across domains. Thus, given the imperfect awareness of LLM hallucinations, establishing such mandates represents a key step in protecting LLM users from their own mistakes.\footnote{\citet{pearson2025internet} suggests that LLM hallucinations may seed false memories and even alter how we remember the past.} In addition, such mandates effectively mitigate domain-specific misinformation externalities, whereas implementing market-based instruments for the same purpose may be impractical. 

Furthermore, I consider the challenge of LLM hallucinations from both developers' and users' perspectives. For instance, LLM developers face a trade-off between minimizing hallucinations and maintaining other model performance metrics \citep{hron2024training}. LLM users, on the other hand, have varying tolerance levels for LLM hallucinations depending on the domain of use, ranging from zero tolerance in critical fields like medicine to greater acceptance in applications like brainstorming. My behavioral welfare decomposition demonstrates that the impact on net welfare from establishing domain-specific maximum hallucination standards is directly linked to both of the preceding considerations. 

This paper builds upon the literature on policy instrument choice under uncertainty \citep[e.g.,][]{ellerman2003absolute,newell2008indexed}. In particular, when examining the welfare economics of fuel economy standards under uncertain future compliance costs, \citet{kellogg2018gasoline} adapts the standard model of \citet{weitzman1974} to show that indexing the fuel economy standard to both the price of gasoline and improvements in fuel economy technology achieves the first-best outcome. Along similar lines, in this paper, I demonstrate that net welfare improves when the maximum acceptable level of LLM hallucinations is designed to vary with two domain-specific factors: the willingness to pay for reduced LLM hallucinations and the marginal damage associated with misinformation.\footnote{Within domain-specific standards, both of these factors can be considered locally constant in the model.}


This work is also related to the literature examining policy designs for behavioral agents in the context of choosing energy-using durables \citep[e.g.,][]{allcott2014energy,gerster2024optimal,houde2019heterogeneous}. I leverage the insight from \citet{houde2019heterogeneous} that under heterogeneous consumer misperceptions, energy efficiency standards reduce the variance of the potentially misperceived attribute within the choice set. In the context of regulating LLM hallucinations, even conditional on the domain of use, the regulator could still face substantial heterogeneity in users' imperfect awareness of LLM hallucinations \citep{bergemann2025economics}. Following \citet{houde2019heterogeneous}, I show that an optimal domain-specific mandate which minimizes the variance in LLM hallucinations allows for the internalization of imperfect awareness while simultaneously addressing the misinformation externality. Therefore, domain-specific standards may be desirable, given that regulators are likely averse to uncertainty about awareness levels of LLM hallucinations within each domain of use.

\section{Model and Analysis}
I consider how an entrepreneur chooses a designated version of a general-purpose LLM created by firms, after which the entrepreneur curates a domain-specific dataset and employs fine-tuning techniques to develop a domain-specific LLM for end users. I begin with an entrepreneur $n$ of domain type $d$ who is considering purchasing a general-purpose LLM, represented by $l$ (i.e., the product). I use the following parsimonious specification to characterize the entrepreneur's utility when choosing a product $l$:
\begin{equation}
U_{n,l,d} = \delta_l - \alpha_d P_l - \theta_d H_l + \varepsilon_{n,l,d}. \label{modelbaseline}
\end{equation}

In Equation (\ref{modelbaseline}), the constant $\delta_l$ represents the overall quality of product $l$. I examine two main product attributes: price $P_l$ and hallucination tendency $H_l$. A higher value of hallucination tendency indicates that the LLM hallucinates more frequently. I let $\alpha_d$ and $\theta_d$ be nonnegative domain-specific scalar parameters, reflecting that utility decreases as purchase price and hallucination tendency increase, with all else held constant. Therefore, the ratio $\theta_d/\alpha_d$ represents the entrepreneur's willingness to pay for reduced LLM hallucinations. 

To capture the possibility that an entrepreneur of type $d$ may not fully account for the disutility associated with LLM hallucinations, I introduce a domain-specific awareness parameter $\rho_d$ and rewrite an entrepreneur's decision utility as
\begin{equation}
U_{n,l,d} = \delta_l - \alpha_d P_l - \theta_d \rho_d H_l + \varepsilon_{n,l,d}. \label{modelmain}
\end{equation}
In Equation (\ref{modelmain}), I assume that $\rho_d \in [0,1].$ Specifically, $\rho_d=0$ if the entrepreneur completely ignores the fact that LLMs are prone to hallucination, while $\rho_d=1$ if the entrepreneur is perfectly aware of such hallucinations.\footnote{However, complete ignorance and perfect awareness are unlikely.}

Applying the standard set of logit model assumptions, I can rewrite Equation (\ref{modelmain}) as $U_{n,l,d} = V_{l,d} + \varepsilon_{n,l,d}$ in which $V_{l,d} = \delta_l - \alpha_d P_l - \theta_d \rho_d H_l$ is the deterministic part of the utility while each $\varepsilon_{n,l,d}$ is independently and identically distributed extreme value. Therefore, the probability of a type $d$ entrepreneur choosing product $l$ is
\begin{equation}
s_{l,d} = \frac{\exp(V_{l,d})}{\sum_{l=1}^{L} \exp(V_{l,d})}. \label{choiceprob}
\end{equation}

When characterizing entrepreneurs' consumer surplus, I follow \citet{leggett2002environmental} in distinguishing between entrepreneurs' \textit{ex ante} decision utility and their \textit{ex post} experienced utility. I make adjustments such that
\begin{align}
CS_d &= \frac{1}{\alpha_d} \left[ \ln \left( \sum^L_{l=1} \exp(V_{l,d}) \right) + \sum^L_{l=1} \left( s_{l,d} \left( \widetilde{V}_{l,d} - V_{l,d} \right) \right) \right], \label{csspec} 
\end{align}
in which $\widetilde{V}_{l,d} = \delta_l - \alpha_d P_l - \theta_d H_l.$

In Equation (\ref{csspec}), $V_{l,d}$ represents the \textit{ex ante} utility considered by a type $d$ entrepreneur when deciding which product $l$ to choose, and $\widetilde{V}_{l,d}$ represents the \textit{ex post} utility experienced by the entrepreneur after putting the product into use. The former incorporates the possibility that the entrepreneur fails to fully account for the disutility associated with LLM hallucinations before using the product while the latter indicates that the original imperfect awareness is eliminated (i.e., $\rho_d=1$) after deploying the product. 

Because $\alpha_d$ represents the marginal utility of income, the consumer surplus expressed in Equation (\ref{csspec}) is the dollar term of the sum of the utility received by the entrepreneur in a choice situation, plus an adjustment based on the entrepreneur's experienced utility.\footnote{In Equation (\ref{csspec}), note that 1) the constant of integration, which reflects the fact that the absolute level of utility cannot be measured, is ignored because it cancels out in subsequent calculations and is therefore irrelevant from a policy perspective; 2) the adjustment term is weighted using choice probabilities computed based on decision utility, and $s_{l,d}$ is a function of $H_l$; 3) $\widetilde{V}_{l,d} - V_{l,d} =  (\rho_d -1) \theta_d H_l \leq 0$ because $\rho_d \in [0,1].$}

To distill intuition, I follow \citet{houde2019heterogeneous} in employing the simplifying assumption that $P_l(H_l) = c(H_l)+\omega_l$, in which $c(H_l)$ is the cost of developing an LLM with a hallucination level of $H_l$, and $\omega_l$ is a product-specific constant markup. Furthermore, I assume that $c'(H_l)<0$ and $c''(H_l)>0$. 

To examine the net welfare from adopting product $l$, I consider a domain-specific misinformation externality arising from LLM hallucinations. Assuming a misinformation externality with constant marginal damage $\zeta_d$, the net welfare is given by
\begin{equation}
NW_d =  CS_d - \zeta_d \sum^L_{l=1} (s_{l,d} H_l). \label{nwspec}
\end{equation}

\subsection{Domain-Specific Hallucination Mandates}
Aiming to maximize the net welfare defined in Equation (\ref{nwspec}), I start with a baseline case, in which the regulator knows every entrepreneur's domain of model use (i.e., $d$), preference (i.e., $\varepsilon_{n,d,l}$), and there is no imperfect awareness (i.e., $\rho_d=1$). With this perfect information, the regulator maximizes consumer surplus for each entrepreneur while accounting for the misinformation externality by solving the following optimization problem:
\begin{align}
\nonumber \max_{H_l} \quad NW_d&=\frac{1}{\alpha_d} V_{l,d} - \zeta_d H_l \\
\nonumber & = \frac{1}{\alpha_d}\delta_l - c(H_l) - \omega_l - \frac{\theta_d}{\alpha_d} H_l - \zeta_d H_l \\
\text{FOC:} \quad -c'(H_l) & = \frac{\theta_d}{\alpha_d} + \zeta_d. \label{perfectinfoH}
\end{align}

Given that one can empirically assess domain-specific willingness to pay for reduced LLM hallucinations and bound the resulting misinformation externality, Equation (\ref{perfectinfoH}) indicates that establishing domain-specific hallucination mandates maximizes net welfare. These mandates would request each version of product $l$ to be tailored to different domain-specific needs, reflecting entrepreneurs' domain-specific willingness to pay for reduced hallucinations and internalizing the corresponding misinformation externality. 

Next, I consider a case without perfect information, in which the regulator only knows there are different domain use types but not each entrepreneur's preference, while domain-specific imperfect awareness exists. In this more general case represented by Equation (\ref{modelmain}), the regulator maximizes net welfare for each domain use type $d$ as follows
\begin{align*}
\max_{H_l,\forall l \in \mathcal{L}} \quad NW_d &= CS_d - \zeta_d \sum^L_{l=1} (s_{l,d} H_l) \\ 
& =\frac{1}{\alpha_d} \left[ \ln \left( \sum^L_{l=1} \exp(V_{l,d}) \right) + \sum^L_{l=1} \left( s_{l,d} \left( \widetilde{V}_{l,d} - V_{l,d} \right) \right) \right] - \zeta_d \sum^L_{l=1} (s_{l,d} H_l).
\end{align*}
The solution to this maximization problem is characterized by a set of FOCs requiring $\frac{\partial NW_d}{\partial H_j}=0$ for each $j \in \mathcal{L}$ in which $\mathcal{L}=\{1,2,\cdots,L\}$ is the set of all LLM products. As shown in Appendix \ref{appoptimalH}, this yields 
\begin{equation}
-c'(H_j)=\frac{\theta_d}{\alpha_d} + \zeta_d + \frac{1}{s_{j,d}} \sum^L_{l=1} \left( \frac{\partial s_{l,d}}{\partial H_j} H_l \right) \left( \zeta_d - (\rho_d - 1) \frac{ \theta_d}{\alpha_d}  \right) \label{generalH}
\end{equation}
Similar to Equation (\ref{perfectinfoH}), net welfare for domain type $d$ is maximized when there is a domain-specific hallucination mandate, which features $\frac{\partial s_{l,d}}{\partial H_j}=0$ in the third term on the right-hand side of Equation (\ref{generalH}). 

\subsection{Welfare Effect of Standards}
The previous section shows that net welfare is maximized under domain-specific hallucination mandates, which require all LLMs for the same domain-use type to stick to the same level of hallucination. This section examines how a more realistic domain-specific maximum hallucination standard improves net welfare relative to a scenario in which there isn't any such standards established. 

First, I put together Equations (\ref{csspec}) and (\ref{nwspec}) to obtain a more explicit expression of net welfare when there isn't any standards as the following 
\begin{equation}
NW_d = \frac{1}{\alpha_d} \left[ \ln \left( \sum^L_{l=1} \exp(V_{l,d}) \right) + (\rho_d - 1)\theta_d \sum^L_{l=1} (s_{l,d}H_l) \right] - \zeta_d \sum^L_{l=1} (s_{l,d} H_l). \label{nw_wostandard}
\end{equation}

Next, I introduce a domain-specific maximum hallucination standard that says all entrepreneurs of domain-use type $d$ should choose LLM products with hallucination rates not larger than a maximum value. When such a standard is in effect, firms update $H_l$ to $\overline{H}_l$, which leads to corresponding $\overline{V}_{l,d}$ and $\overline{s}_{l,d}.$ Then the net welfare under this new standard becomes
\begin{equation}
\overline{NW}_d = \frac{1}{\alpha_d} \left[ \ln \left( \sum^L_{l=1} \exp(\overline{V}_{l,d}) \right) + (\rho_d - 1)\theta_d \sum^L_{l=1} (\bar{s}_{l,d}\overline{H}_l) \right] - \zeta_d \sum^L_{l=1} (\bar{s}_{l,d} \overline{H}_l). \label{nw_wstandard}
\end{equation}

Comparing Equations (\ref{nw_wostandard}) and (\ref{nw_wstandard}) allows me to decompose changes in net welfare into three components as shown below
\begin{align*}
\Delta NW_d &= \overline{NW}_d - NW_d \\
&= \underbrace{\frac{1}{\alpha_d} \ln \left( \frac{\sum^L_{l=1} \exp(\overline{V}_{l,d})}{\sum^L_{l=1} \exp(V_{l,d})} \right)}_{I} \\
& \quad + \underbrace{(\rho_d - 1) \frac{\theta_d}{\alpha_d} \left[ \sum^L_{l=1} (\bar{s}_{l,d}\overline{H}_l) - \sum^L_{l=1} (s_{l,d}H_l) \right]}_{II} \quad \underbrace{- \quad \zeta_d \left[ \sum^L_{l=1} (\bar{s}_{l,d}\overline{H}_l) - \sum^L_{l=1} (s_{l,d}H_l) \right]}_{III}.
\end{align*}

A maximum hallucination standard improves welfare by excluding products with hallucination levels exceeding the specified threshold. When such a standard is established, the choice probability weighted average hallucination rate will be no greater than its value without the standard. Therefore, the difference characterized by $\left[ \sum^L_{l=1} (\bar{s}_{l,d}\overline{H}_l) - \sum^L_{l=1} (s_{l,d}H_l) \right]$ in Components II and III of $\Delta NW_d$ will be nonpositive.

To examine the welfare effect of a standard, I begin with Component II. Given that $\rho_d \in [0,1]$ and $\left[ \sum^L_{l=1} (\bar{s}_{l,d}\overline{H}_l) - \sum^L_{l=1} (s_{l,d}H_l) \right] \leq 0$, Component II will be nonnegative. The magnitude of Component II depends on entrepreneurs' willingness to pay for reduced LLM hallucinations, scaled by the gap between the actual awareness level of LLM hallucinations and perfect awareness. The former term indicates that in domains such as medical diagnosis, which demand absolute factual accuracy and feature a high willingness to pay for reduced LLM hallucinations, the gain from establishing a maximum hallucination standard could be substantial. The latter term reflects the gap between the decision and experienced utility, which is considerable for entrepreneurs who are the least aware of LLM hallucinations. 

Meanwhile, Component III shows that, as long as the misinformation damage exists (i.e., $\zeta_d>0$), a maximum hallucination standard that moves products with high hallucination levels in the choice set to become compliant with the standard, always offers welfare improvement by mitigating misinformation externalities.

The sign of Component I in $\Delta NW_d$ hinges on the relative magnitude of $\sum^L_{l=1} \exp(\overline{V}_{l,d})$ and $\sum^L_{l=1} \exp(V_{l,d})$. This relationship depends on the domain-specific willingness to pay for reduced LLM hallucinations and the corresponding imperfect awareness of those hallucinations. Conditional on domain type, technological progress, and firms' strategic interactions and adaptations, it is possible to derive a maximum acceptable level of LLM hallucinations that yields a positive perceived net private benefit.\footnote{Regulatory mandates can influence technological advancements. For example, \citet{rozendaal2021policy} demonstrate that, in the passenger car sector, standards for greenhouse gas emissions and fuel economy induce innovations in clean car technologies. Similarly, in the case of smart grid technology, \citet{gregoire2024technology} provide suggestive evidence showing that interoperability standards improve innovation quality. However, firms' strategic interactions and adaptations are more complex, as illustrated by recent investigations into PC makers and car manufacturers \citep[e.g.,][]{eizenberg2014upstream,ito2018economics,reynaert2021abatement,reynaert2021benefits}. These investigations highlight the uncertainty regarding empirical parameters related to regulating LLM hallucinations and point to areas for future research.}

\section{Concluding Remarks}
LLMs are tools that promise great productivity gains. However, they come with potential pitfalls, such as data privacy issues, copyright infringement, and environmental harm from powering and cooling the data centers supporting them. This paper examines LLM hallucinations. It shows that, in the presence of imperfect awareness of LLM hallucinations and associated misinformation externalities, establishing maximum hallucination standards improves welfare. Furthermore, the domain of use determines both the acceptable level of LLM hallucinations and the marginal damage from misinformation. Therefore, domain-specific standards help address regulatory needs and balance the trade-off between minimizing hallucinations and maintaining other model performance metrics.

\bibliographystyle{chicago} 
\bibliography{LLMStandards}

\newpage

\begin{appendices}
\onehalfspacing
\addappheadtotoc 
\appendixpage
    


\section{Optimal Mandates \label{appoptimalH}}
Consider the decision utility with imperfect awareness and corresponding net welfare:
\begin{align*}
U_{n,l,d} &= \delta_l - \alpha_d P_l - \theta_d \rho_d H_l + \varepsilon_{n,l,d} \\
\max_{H_l,\forall l} \quad NW_d &= CS_d - \zeta_d \sum^L_{l=1} (s_{l,d} H_l) \\ 
& =\frac{1}{\alpha_d} \left[ \ln \left( \sum^L_{l=1} \exp(V_{l,d}) \right) + \sum^L_{l=1} \left( s_{l,d} \left( \widetilde{V}_{l,d} - V_{l,d} \right) \right) \right] - \zeta_d \sum^L_{l=1} (s_{l,d} H_l)\\
& =\frac{1}{\alpha_d} \left[ \ln \left( \sum^L_{l=1} \exp(V_{l,d}) \right) + \sum^L_{l=1} \left(s_{l,d}(\rho_d - 1)\theta_d H_l \right) \right]- \zeta_d \sum^L_{l=1} (s_{l,d} H_l)\\
& = \frac{1}{\alpha_d} \left[ \ln \left( \sum^L_{l=1} \exp(V_{l,d}) \right) + (\rho_d - 1)\theta_d  \sum^L_{l=1}(s_{l,d}H_l) \right]- \zeta_d \sum^L_{l=1} (s_{l,d} H_l).
\end{align*}
The solution to this maximization problem requires $\frac{\partial NW_d}{\partial H_j}=0$ for each $j \in \mathcal{L}$.

Let $X = \ln \left( \sum^L_{l=1} \exp(V_{l,d}) \right)$, $Y = \sum^L_{l=1} (s_{l,d} H_l)$, and $Z= \zeta_d \sum^L_{l=1} (s_{l,d} H_l)$. Given that 
\begin{align*}
\frac{\partial V_{j,d}}{\partial H_j} &= -\alpha_d c'(H_j) - \theta_d \rho_d \quad \text{and} \\
\frac{\partial X}{\partial V_{j,d}} &= \exp(V_{j,d}) \frac{1}{\sum^L_{l=1} \exp(V_{l,d})} = s_{j,d},
\end{align*}
I obtain
\begin{align*}
\frac{\partial X}{\partial H_j} & = \frac{\partial X}{\partial V_{j,d}} \frac{\partial V_{j,d}}{\partial H_j} \\
&= \left(-\alpha_d c'(H_j) - \theta_d \rho_d \right) s_{j,d}.   
\end{align*}

In the meantime, 
\begin{align*}
\frac{\partial Y}{\partial H_j} &= \left(\frac{\partial s_{1,d}}{\partial H_j}H_1 + 0\right) + \left(\frac{\partial s_{2,d}}{\partial H_j}H_2 + 0\right) + \dots \\
& \qquad + \left(\frac{\partial s_{j,d}}{\partial H_j}H_j + s_{j,d}\right) + \dots + \left(\frac{\partial s_{L,d}}{\partial H_j}H_L + 0\right) \\
&= s_{j,d} + \sum^L_{l=1} \left( \frac{\partial s_{l,d}}{\partial H_j} H_l \right).
\end{align*}

Additionally,
\begin{align*}
\frac{\partial Z}{\partial H_j} &= \frac{\partial \left( \zeta_d \sum^L_{l=1} (s_{l,d} H_l)\right)}{\partial H_j} \\
&= \zeta_d s_{j,d} +  \zeta_d \sum^L_{l=1} \left( \frac{\partial s_{l,d}}{\partial H_j} H_l \right).
\end{align*}

\begin{align*}
\frac{\partial NW_d}{\partial H_j} &= \frac{1}{\alpha_d} \left[ \left( -\alpha_d c'(H_j) - \theta_d \rho_d \right) s_{j,d} + (\rho_d - 1) \theta_d \left(s_{j,d} + \sum^L_{l=1} \left( \frac{\partial s_{l,d}}{\partial H_j} H_l \right) \right) \right] \\
& \qquad - \left( \zeta_d s_{j,d} +  \zeta_d \sum^L_{l=1} \left( \frac{\partial s_{l,d}}{\partial H_j} H_l \right) \right) \\
&= \frac{1}{\alpha_d} \left[ -\alpha_d c'(H_j)s_{j,d} - \theta_d s_{j,d} + (\rho_d - 1)\theta_d \sum^L_{l=1} H_l \left( \frac{\partial s_{l,d}}{\partial H_j} H_l \right) \right]  \\
& \qquad - \left( \zeta_d s_{j,d} +  \zeta_d \sum^L_{l=1} \left( \frac{\partial s_{l,d}}{\partial H_j} H_l \right) \right) \\
&= -c'(H_j)s_{j,d} - \frac{\theta_d}{\alpha_d}s_{j,d} + (\rho_d - 1) \frac{\theta_d}{\alpha_d} \sum^L_{l=1} \left( \frac{\partial s_{l,d}}{\partial H_j} H_l \right) - \zeta_d s_{j,d} - \zeta_d \sum^L_{l=1} \left( \frac{\partial s_{l,d}}{\partial H_j} H_l \right)
\end{align*}

\begin{align*}
\text{FOC:} \quad -c'(H_j) &=  \frac{\theta_d}{\alpha_d} -  \frac{1}{s_{j,d}} (\rho_d - 1) \frac{ \theta_d}{\alpha_d}\sum^L_{l=1}\left( \frac{\partial s_{l,d}}{\partial H_j} H_l \right)  +\zeta_d + \frac{\zeta_d}{s_{j,d}} \sum^L_{l=1} \left( \frac{\partial s_{l,d}}{\partial H_j} H_l \right) \\
&=\frac{\theta_d}{\alpha_d} + \zeta_d + \frac{1}{s_{j,d}} \sum^L_{l=1} \left( \frac{\partial s_{l,d}}{\partial H_j} H_l \right) \left( \zeta_d - (\rho_d - 1) \frac{ \theta_d}{\alpha_d}  \right)
\end{align*}

\qed
\end{appendices}

\end{document}